\documentstyle[aps,prb,psfig,graphicx]{revtex}
\long\def\title#1{{\Large\begin{center}#1\end{center}\par}}
\long\def\address#1{\begin{center}#1\end{center}\par}
\long\def\author#1{\begin{center}#1\end{center}\par}
\def\pacs{}
\tightenlines
\begin{document}

\title{Analytical derivation of lateral pressure profile from microscopic model 
of lipid bilayer}

\author{Sergei I. Mukhin and Svetlana Baoukina}
\address{Theoretical Physics Department, Moscow Institute for Steel and Alloys
4 Leninsky pr., 119049 Moscow Russia\\
e-mail: sergeimoscow@online.ru}

\vspace{3mm}

\begin{abstract}
Field-theoretical method is proposed, that yields analytic expression for 
lateral pressure distribution across hydrophobic core of a bilayer lipid membrane.
Lipid molecule, repelling entropicaly surrounding  neighbors, is modeled as a 
flexible string embedded  between confining walls in the self-consistent harmonic
potential. At room temperature the pressure profile is expressed solely via 
lowest energy eigenfunctions of the operator of the string energy density. 
Analytical results compare favorably with molecular dynamics simulations. 
The theory is suited to study protein-lipid interaction.
\end{abstract}

\vspace{3mm}


\vspace{3mm}

\begin{figure}
 \vbox to 4.0cm {\vss\hbox to -5.0cm
 {\hss\
       {\includegraphics{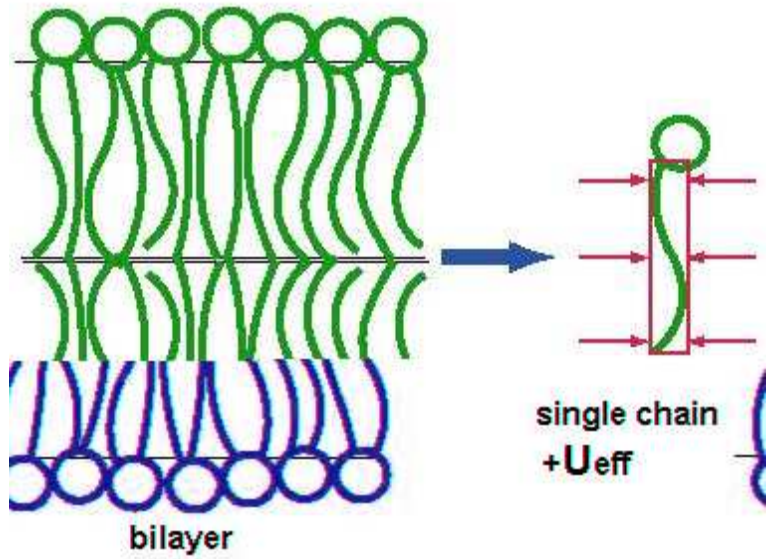}
       }
  \hss}
 }
\vspace{12.0cm}
\caption{Mean-field approximation for inter-chain interactions in lipid bilayer.}
\label{fig1}
\end{figure}

\begin{figure}
 \vbox to 4.0cm {\vss\hbox to -5.0cm
 {\hss\
       {\includegraphics{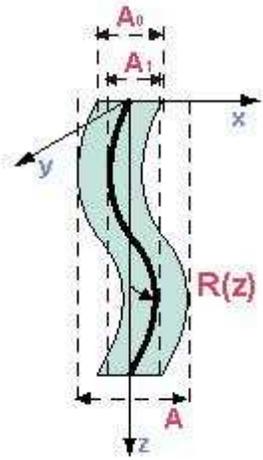}
       }
  \hss}
 }
\vspace{12.0cm}
\caption{Hydrocarbon chain as a flexible string of finite length. 
$\vec{R}$ is vectorin $\{x,y\}$ characterizing deviation of the chain 
from z-axis; $A=\pi<\vec{R}^2>\equiv \pi d^2$ is average area per string.}
\label{fig2}
\end{figure}

Development of microscopic theory of fluid lipid membrane is a nontrivial problem.
One of the fundamental difficulties lies in the elongated flexible structure of 
the constituting lipid molecules that results in their large conformational 
entropy. Neighboring lipids essentially reduces the manifold of accessible 
conformations of the molecule's hydrocarbon chains in a bilayer membrane as 
compared with a free lipid molecule. This reduction, in turn, leads to effective
 entropic repulsion between lipid molecules of non-local nature. Entropic 
interaction \cite{Helfrich} between hydrocarbon chains causes unusual for 
common fluids inhomogeneous lateral pressure distribution in thermodynamic 
equilibrium state. The lateral pressure profile strongly influences the 
functioning of proteins embedded in the membrane \cite{Cantor}, \cite{Marsh}, 
for instance by shifting their activation barriers 
\cite{Perozo}, \cite{Sukharev}. 
Due to a nanometer scale of membrane thickness the pressure profile is difficult
 to measure. Several attempts to attack the present problem theoretically have
 already been made. Some of them involved molecular dynamics simulations of a 
lipid bilayer \cite{Lindahl}, \cite{Gullingsrud}. 
The others exploited numerical methods (e.g. Monte Carlo) with mean-field 
approach, introducing lateral pressure profile as a substitute for
 inter-molecular entropic repulsion \cite{Cantor},\cite{Harries},\cite{Ben-Shaul}. 
In the present work we use path-integral technique in combination with adiabatic
 invariants theorem \cite{Landau} and find lateral pressure profile in 
analytical form. Due to its generality our method permits, in principle, 
analytical derivation of the lateral pressure profile for a great variety 
of membrane constituents and lipid variations (head group type, chain length, 
etc.) by appropriate change of the imposed boundary conditions and membrane 
spontaneous curvature. Hence, our theory, besides fundamental interest, may 
be used as a tool for theoretical prediction of the properties of a wide 
variety of lipid membranes.
A sketch of lipid bilayer membrane is given in Fig.\ref{fig1}. Hydrocarbon chains 
form hydrophobic part in lipid molecules. Collisions between chains together 
with excluded volume effect give rise to entropic repulsion between lipid 
molecules. This repulsion is responsible for inhomogeneous lateral pressure 
in the hydrophobic core of the membrane.
The ensemble of chains is described within mean-field approximation using 
microscopic model of a single hydrocarbon chain in effective potential, see 
\ref{fig1}. The chain is modeled as a flexible string (see Fig. \ref{fig2}) 
that possesses bending rigidity, $K_f$. 
The effective potential, which allows for entropic repulsion between neighboring 
chains in each monolayer is expressed via lateral pressure distribution $\Pi(z)$
 at the depth $z$ multiplied by area $\pi R^2(z)$ swept by the chain in $\{x,y\}$
-plane:$U_{eff}=\pi \Pi R^2(z)$ (compare e.g. 
\cite{Ben-Shaul}). Here the chain deviation from the $z$-axis is 
$R^2=R^2_x+R^2_y$. The energy functional of a single hydrocarbon chain $E_t$  is 
chosen in the following form:

\begin{eqnarray}
E_t=\displaystyle{\int^L}_0\left\{\frac{K_f}{2}\left[\frac{\partial^2R(z)}{\partial z^2}
\right]^2+\pi \Pi(z)R^2(z)\right\}dz.
\label{et}
\end{eqnarray}

\noindent
Here $L$ is chain length; $z$ is coordinate along the chain axis; $\vec{R}$is vector 
in $\{x,y\}$ plane characterizing chain deviation from $z$-axis.
In the functional (\ref{et}) the bending energy term contains second derivative 
over $z$-coordinate rather than over the contour length of the chain. The approximation 
made is valid provided that chain deviations from the $z$-axis are small with respect 
to chain length $L$ (applicability of this condition at room temperature is justified 
below). We rewrite the energy functional $E_t$ using self-adjoint operator,$\hat{H}$ , 
that is obtained after integrating by parts the expression in (\ref{et}): 

\begin{eqnarray}
E_t=\displaystyle\sum_{i=x,y}\int^L_0 R_i(z)\hat{H}R_i(z)dz\;;\hat{H}=
\displaystyle
\frac{1}{2}K_f\frac{\partial^4}{\partial z^4}+\pi \Pi(z).
\label{eth}
\end{eqnarray}

The boundary conditions are: i) chain angle is fixed in the head group region; 
ii) zero torque is applied at the free chain end; iii) no total force is applied 
to the chain at head group and at the free chain end. These conditions are expressed 
as: $R'(0)=R'''(0)=R''(L)=R'''(L)=0$. An arbitrary chain conformation described by 
"shape-functions" $R_{x,y}(z)$ can be expanded over orthonormal set of the 
eigenfunctions, $R_n(z)$, of operator $\hat{H}$:
	
\begin{eqnarray}
R_x=\sum C_nR_n(z).
\label{eigs}
\end{eqnarray}

\noindent
Since $x$- and $y$-deviations are independent we shall retain for simplicity only one 
index, e.g.$x$. Then, the energy functional $E_t$ of hydrocarbon chains given in 
(\ref{eth}) takes equivalent form:

\begin{eqnarray}
E_t=\int^L_0\sum_n E_nC_n^2R_n^2(z)dz,
\label{enhe}
\end{eqnarray}
 
\noindent
where $E_n$ are the eigenvalues of the operator $\hat{H}$ in (\ref{eth}). 

To find the distribution of lateral pressure along $z$-axis we shall use the following 
formal trick \cite{Mukhin}. Imagine that we press the membrane laterally at depth 
$z$ and measure its response. This would give us local lateral pressure. Mathematically 
this procedure can be performed using the adiabatic invariants theorem 
\cite{Landau}. Consider the energy (ref{enhe}) as a sum of potential 
energies of harmonic oscillators labeled by index $n$ and coordinate $z$ that have 
rigidities $k_n=2E_n$  and vibration amplitudes $X_n=C_nR_n(z)$  \cite{multiplier}. 
Then, the lateral pressure is a sum of individual contributions produced by the oscillators
defined above. We calculate each contribution as a thermodynamically averaged force exerted 
by a one-dimensional oscillator on the surrounding "walls". Assume that the oscillator is 
centered at $x=0$ and the walls are positioned at points $\pm d(t)$, where $d(t)=d-vt$ and 
$v\rightarrow 0$. Here $d$ is $z$-independent radius of a cylinder, which on average hosts 
a single chain in a flat bilayer, see Fig.\ref{fig2}. The rate of the energy change 
$d\epsilon_{nz}/dt$ of n-th 
oscillator at depth $z$ is related to the force, $f_{n,z}$, exerted by the oscillator on the 
walls:$d\epsilon_{nz}/dt=2vf_{n,z}$. Alternatively, the rate of the energy change can be 
calculated using the method of adiabatic invariants \cite{Landau}:

\begin{eqnarray}
d\epsilon_{nz}/dt=v\displaystyle\frac{\partial I/\partial d}{\partial I/\partial \epsilon_{n,z}}
\label{adi}
\end{eqnarray}
 	
where time-independent adiabatic invariant is :$I={\int} pdx$. Here
$p=\sqrt{2m(\epsilon_{n,z}-k_nx^2/2)}$ is momentum, $x$ is coordinate of the 
oscillator (mass $m$ of the oscillator does not enter the final expression).
Integration in $I$ gives area inside  the closed orbit in phase-space $\{p,x\}$ 
crossing at points $\pm d(t)$  the $x$-axis and at $\pm p(d(t))$ the p-axis. 
Equating the above two expressions for $d\epsilon_{nz}/dt$ and taking limit 
$v\rightarrow 0$ we find the force:

\begin{eqnarray}
f_{n,z}=\left\{\begin{array}{c}
\displaystyle\frac{\sqrt{k_n\epsilon_{nz}}\sqrt{1-d^2k_n/2\epsilon_{nz}}}
{\sqrt{2}\arcsin(d\sqrt{k_n/2\epsilon_{nz}})},\;d^2k_n/2\epsilon_{nz}<1,\\
0,\;d^2k_n/2\epsilon_{nz}>1.
\end{array}\right.
\label{f}
\end{eqnarray}

The force produced by n-th oscillator at depth $z$ is found as a result of 
averaging over $C_n$:

\begin{eqnarray}
<f_n(z)>=\int^{\infty}_{-\infty}f_n(\epsilon_{nz}(C_n))P(C_n)dC_n,
\label{fm}
\end{eqnarray}

\noindent
where $\epsilon_{nz}(C_n)=E_nC_n^2R_n^2(z)$ and the distribution function equals:

\begin{eqnarray}	
P(C_n)=\displaystyle\left(E_n/\pi k_BT\right)^{1/2}\exp\left\{-\frac{1}{k_BT}
\int^L_0 E_nC_n^2R_n^2(z)dz \right\}.
\label{cn}
\end{eqnarray}

Summation over n permits to find the total force at depth $z$:
	
\begin{eqnarray}
<f(z)>=\displaystyle\sum_n\frac{1}{\sqrt{\pi k_BTR_n^2(z)}}\int^{\infty}_{0}
\frac{f_n(\epsilon)}{\sqrt{\epsilon}}\exp\left\{-\frac{\epsilon}{R_n^2(z)k_BT}
\right\}d\epsilon ,
\label{fmt}
\end{eqnarray}

Obtained formula is remarkable: though probabilities for "oscillators" $n$ with 
different $z$'s are distributed independently according to formula (\ref{cn}), 
nevertheless, factors $R_n(z)$ in formula (\ref{fmt}) keep track of the fact that 
"oscillators" are different degrees of freedom of one and the same vibrating string. 
Actually, only the oscillators that reach the walls at $\pm d$ contribute to the force 
$f(z)$. This bounds energy $\epsilon$ from below:$\epsilon\geq d^2k_n/2$. Hence, the 
lower limit of integration in (\ref{fmt}) increases fast with index n, because bending 
part of total energy in (\ref{et}) results in power-law dependence: $k_n=2E_n\propto n^4$. 
Simultaneously, there is an upper limit of integration in (\ref{fmt}) $\sim R_n^2(z)k_BT$
imposed by the Boltzmann factor. Hence, non-zero integration interval exists only for the 
first few terms in (\ref{fmt}). Here we limit ourselves to the terms $n=0,1$ and find the 
thermodynamically averaged force:

\begin{eqnarray}
<f(z)>\approx\displaystyle\sum_n\frac{k_BT}{2d}R_n^2(z).
\label{fmta}
\end{eqnarray}

\noindent
As the membrane is laterally isotropic, this force can be considered as a radial force 
in the $\{x-y\}$ plane. Then, the pressure distribution is obtained dividing the radial 
force in (\ref{fmta}) by half the perimeter $\pi d$ :

\begin{eqnarray}
\Pi(z)\approx\displaystyle\sum_n\frac{k_BT}{2\pi d^2}R_n^2(z).
\label{fin}
\end{eqnarray}

This is the central result of our paper, since, we had shown that the lateral pressure 
profile that enters the chain energy functional $E_t$ in Eq. (\ref{et}) is expressed 
merely via the eigenfunctions of the energy density-operator $\hat{H}$. Using this result
we find the corresponding self-consistent solution for $\Pi(z)$  analytically at room 
temperature. From the estimates for the chain bending rigidity: 
$K_f\sim k_bTl_p$, where $l_p$ is chain persistence length, it proves that the separation 
between the eigenvalues $E_n$ of the operator $\hat{H}$ is large with respect to $k_BT$. 
This justifies approximate projection of the unknown eigenfunctions $R_n(z)$ on the 
reduced basis formed by the lowest energy eigenfunctions of the free-chain operator
$\hat{H}_0=K_f/2(\partial^2/\partial z)$:  

\begin{eqnarray}
R_0=a_1Y_0+b_1Y_1;\;R_1=a_2Y_0+b_2Y_1,
\label{eif}
\end{eqnarray}

\noindent	
where:

\begin{eqnarray}
Y_0(z)=1/\sqrt{L},\;Y_{n\geq 1}(z)=\sqrt{2/L}\{\cos(\lambda_nz)+\cos(\lambda_nL)
\cosh(\lambda_nz)/\cosh(\lambda_nL)\} ,\nonumber
\end{eqnarray}

\noindent
and the eigenvalues are given by:

\begin{eqnarray}
W_0=0,\;
W_{n\geq 1}=K_f/(2L^4)(\pi/4)^4(4n+3)^4,
\nonumber
\end{eqnarray}

\noindent
where $\lambda_{n\geq 1}L=3\pi/4+\pi n$. The coefficients $a_i$, $b_i$ satisfy the 
conditions of orthonormality of the functions $R_n(z)$: 

\begin{eqnarray}
a_i^2+b_i^2=1,\; a_1b_1+a_2b_2=0
.\nonumber
\end{eqnarray}

To find the lowest energy eigenfunctions and eigenvalues of the energy-density operator  
in (\ref{eth}) we substitute unknown functions $R_n(z)$ in the form (\ref{eif}) into 
expression for $\Pi(z)$  in (\ref{fin}), and insert the resulting expression for $\Pi(z)$
into the equation for eigenfunctions of the operator $\hat{H}$ in (\ref{eth}). We arrive 
at the following system of non-linear algebraic equations for the lowest eigenvalues 
$E_n$ of the operator $\hat{H}$  and coefficients $a_i$, $b_i$:

\begin{eqnarray}
\left\{\begin{array}{c}
\displaystyle a_1\Pi_{00}+b_1\Pi_{01}=a_1E_0\\
\displaystyle b_1 W_{1}+b_1\Pi_{11}+a_1\Pi_{10}=b_1E_0\\
\displaystyle a_2\Pi_{00}+b_2\Pi_{01}=a_2E_1\\
\displaystyle b_2 W_{1}+b_2\Pi_{11}+a_2\Pi_{10}=b_2E_1 
\end{array}\right.
\label{eqs}
\end{eqnarray}

\noindent
where the matrix elements $\Pi_{mn}$  of the "potential energy" operator $\pi \Pi(z)$  
are taken on the set of eigenfunctions $Y_n(z)$  defined above. It proves to be that when
$\Pi_{00}/W_1\leq 10$  the system of equations (\ref{eqs}) can be solved in closed form. 
The solutions are:

\begin{eqnarray}
&&E_0=\Pi_{00}(1-I_3\delta),\; E_1=W_1+\Pi_{00}(1+I_4+I_3\delta),\\
&&(a_1;b_1)=\left(1+\delta^2\right)^{-1/2}(1;-\delta),\;
(a_2;b_2)=\left(1+\delta^2\right)^{-1/2}(\delta;1).
\label{sols}
\end{eqnarray}

\noindent
Here, the following parameters have been introduced:

\begin{eqnarray}
&&I_3=L^{3/2}\int^L_0Y^3_1(z)dz\approx -0.4,\; I_4=L^2\int^L_0Y^4_1(z)dz
\approx 1.8, \nonumber\\\
&&\delta\equiv I_3\Pi_{00}/(W_1+\Pi_{00}I_4)\approx -0.17.
\label{params}
\end{eqnarray}

\noindent
The matrix element $\Pi_{00}$ obeys normalization condition:

\begin{eqnarray}
\Pi_{00}=\displaystyle\frac{1}{L}\int^L_0\pi\Pi(z)dz=\frac{\pi P_{eff}}{L}.
\nonumber
\end{eqnarray}

\noindent
Effective surface tension in the bilayer,$P_{eff}$, balances entropic repulsion between 
hydrocarbon chains. Substituting (\ref{sols}) into (\ref{eif}) and subsequently 
(\ref{eif}) into (\ref{fin}) we find analytical expression for the lateral pressure 
profile:
	
\begin{eqnarray}
\Pi(z)=\displaystyle\frac{P_{eff}}{2(1+\delta^2)}\left\{(Y_0(z)-\delta Y_1(z))^2+
(Y_1(z)+\delta Y_0(z))^2\right\}.
\label{lat}
\end{eqnarray}

\begin{figure}
 \vbox to 4.0cm {\vss\hbox to -5.0cm
 {\hss\
       {\includegraphics{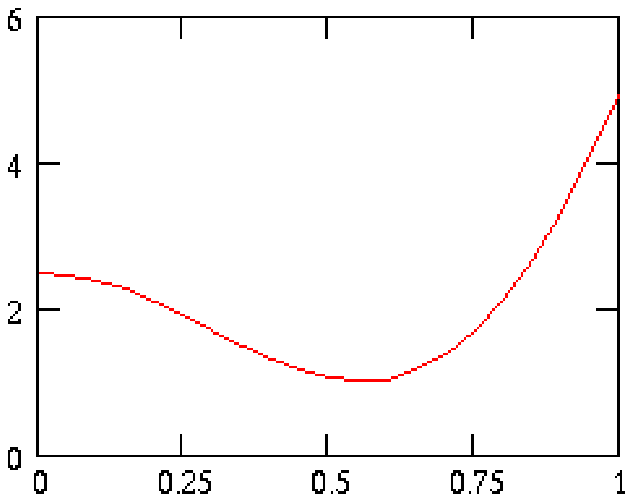}
       }
  \hss}
 }
\vspace{12.0cm}
\caption{Calculated lateral pressure distribution in the hydrophobic core of the 
bilayer. The pressure arises from entropic repulsion between hydrocarbon chains 
in each monolayer. $z$ is coordinate along chain axis spanning from $z=0$ at head 
group to $z=L$ at free chain end. Pressure is normalized by $\Pi_0=P_{eff}/2L$.}
\label{fig3}
\end{figure}

\noindent
Normalized lateral pressure profile (\ref{lat}) is plotted in Fig.\ref{fig3}. 
The system parameters are: chain full length $L=20 A$, area per chain in all-trans 
conformation $A=20 A^2$, chain bending rigidity $K_f=k_BTl_p=2.8\cdot 10^{-5}$ 
erg$\cdot$cm, $T=300K$, $k_B$- Boltzmann constant, $l_p\sim L/3$ - chain persistence 
length,$P_{eff}\approx 150$dyne/cm  \cite{Lindahl}. 
The calculated profile, shown in Fig.\ref{fig3}, qualitatively agrees with the results
of the recent molecular dynamics simulations \cite{Lindahl}, \cite{Gullingsrud}. 
The behaviour of the $\Pi(z)$ curve can be understood comparing 
fluctuations of a lipid chain in bilayer and in empty space. In the bilayer the chain 
fluctuations are strongly suppressed by the neighboring chains with respect to a free 
lipid molecule. Suppression of chain fluctuation leads to the decrease of 
conformational entropy and, in turn, results in the enhancement of entropic pressure. 
The more suppressed the fluctuations, the higher the entropic pressure. The most 
significant relative decrease of fluctuation amplitude takes place at chain free end 
($z\sim L$); at head group region ($z\sim 0$) the relative decrease is noticeable too 
as lipid heads are not fixed; at mid-chain region the decrease of fluctuations is less
pronounced due to segment connectivity in the chain. 
Now we can validate the exploited approximation of small chain deviations in the 
bilayer. We calculate the thermodynamic average of chain fluctuation amplitude 
using expansion (\ref{eigs}) and distribution function (\ref{cn}):

\begin{eqnarray}
<R^2(z)>=2<R^2_{x,y}(z)>=k_BT\sum_n\frac{R^2_n(z)}{E_n}.
\label{eva}
\end{eqnarray}

\noindent
Since,$E_n\propto n^4$, the sum in (\ref{eva}) converges fast. Allowing for relations
$R^2_0(z)\sim 1/L$,$E_0\sim \Pi_{00}\sim P_{eff}/L$, we find rough estimates for the 
small parameter: 

\begin{eqnarray}
\sqrt{<R^2(z)>}/L\sim (k_BT/L^2P_{eff})^{1/2}\sim 0.1.
\label{small}
\end{eqnarray}

To summarize, we have demonstrated that lateral pressure profile in the hydrophobic 
core of lipid bilayer is expressed via the eigenfunctions of the density operator of 
the chain energy. The self-consistent solution for the lateral pressure distribution
is found.

The authors acknowledge valuable discussions with  Prof.Yu. Chizmadzhev and useful 
comments at theoretical seminar at Institute Lorentz . The work of S.I.M. was in part 
supported by RFBR grant 02-02-16354. The work of  S.B. was supported by the Foundation
"Dynasty".

\newpage

\newpage

\newpage

\begin{references}
\bibitem{multiplier} A dimensional multiplier of a unit of length is omitted in front of 
$E_n$ above. Formally, it should be introduced by the transition from the integration 
over $z$ to the summation. Nevertheless, this multiplier merely leads to the 
redefinition of the force calculated in (\ref{fmta})  as being rather a force per unit 
of length.
\end{references}
\end{document}